\documentclass[12pt,a4paaper,oneside]{article}
\usepackage{lineno,hyperref}

\usepackage[margin=1in]{geometry}  
\usepackage{graphicx}              
\usepackage{epsfig}
\usepackage{amsmath}               
\usepackage{amssymb}
\usepackage{epstopdf}
\usepackage{verbatim}
\usepackage{mathptmx}
\usepackage{mathalpha}
\usepackage{mathtools}
\usepackage{graphicx, times}
\usepackage{amsfonts}              
\usepackage{amsthm}                
\usepackage{multicol}
\usepackage{algorithm}
\usepackage{algorithmic}
\usepackage{varwidth}
\usepackage{parskip}
\usepackage{hyperref}
\usepackage{rotating}
\usepackage[numbers,sort&compress]{natbib}
\usepackage{multirow}
\usepackage{pdflscape}
\usepackage[numbers]{natbib}
\usepackage{caption}
\usepackage{xcolor}
\usepackage{color}
\usepackage{comment}
\usepackage{subcaption}
\newcommand*{\rom}[1]{\expandafter\@\romannumeral #1}

\newcommand{\bea}{\begin{eqnarray}}
	\newcommand{\eea}{\end{eqnarray}}
\newcommand{\bee}{\begin{eqnarray*}}
	\newcommand{\eee}{\end{eqnarray*}}

\begin{document}
\author{ Romanshu Garg$^{1}$\footnote{romanshugarg18@gmail.com}, G. P. Singh$^{1}$\footnote{gpsingh@mth.vnit.ac.in}, Ashwini R Lalke$^{2}$\footnote{lalkeashwini@gmail.com}, Saibal Ray$^{3}$\footnote{saibal.ray@gla.ac.in}
\vspace{.3cm}\\
${}^{1}$ Department of Mathematics,\\ Visvesvaraya National Institute of Technology, Nagpur, India.
\vspace{.3cm}
\\${}^{2}$ Departments of Applied Mathematics , \\ Shri Ramdeobaba College of Engineering and
Management, Nagpur, India.
\vspace{.3cm}\\
${}^{3}$ Centre for Cosmology, Astrophysics and Space Science (CCASS),\\
GLA University, Mathura 281406, Uttar Pradesh, India}
\date{}

\title{Cosmological model with linear equation of state parameter in $f(R, L_{m})$ gravity}

\maketitle

\begin{abstract}
In this paper, we examine the universe's expansion in $ f(R, L_{m}) $ gravity for a particular form of $ f(R, L_{m})=\frac{R}{2}+L_{m}^{n}$. The field equations for flat FLRW metric with matter Lagrangian $ L_{m}=\rho$ are derive. Hubble parameter in terms of red-shift$(z)$ are derived using the linear form of Equation of State (EoS) parameter $ \omega=w_{0}+w_{1}z $. By using Bayesian statistical techniques based on $ \chi^{2}$-minimization technique, we have obtained the best fit values of the model parameters of this model for cosmic chronometer and supernovae Pantheon datasets. The evolution of the equation of state parameter$(\omega)$, energy density $ (\rho) $, pressure $\mathit{(p)}$ cosmographic parameters, and the impact of the energy conditions with best-fit values of the model parameters are all thoroughly examined. We have also analyze Om diagnostic's  behavior and determine the present age of universe for this model.
\end{abstract}

\section{Introduction}\label{sec:1}

Astronomical observations provide evidence that the  expansion of the universe is growing~\cite{1998AJ....116.1009R,1999ApJ...517..565P,2020A&A...641A...6P}. Subsequently, researchers in theoretical cosmology developed a desire to construct cosmological models representing an accelerating expansion phase of the universe. To obtain such type of model, one has to either modify Einstein's field equations in the form of the new  form of  alternative theories of gravity or modify the four-dimensional space-time geometry. The exact form of the mysterious candidate responsible for accelerating the expansion of the universe is not yet known and hence referred by the majority of researchers  as dark energy. The various forms of dark energy generating negative pressure have been considered to find a more suitable cosmological model to elucidate the present growth  in the expansion of the universe. It is widely accepted that one of the most promising candidates for dark energy is the dynamical cosmological term $ \Lambda $~\cite{weinberg1989cosmological,ray2004accelerating,usmani2008dark,mukhopadhyay2008lambda,ray2009scenario, mukhopadhyay2009generalized,mukhopadhyay2011time,ray2013scenario}. However, $ \Lambda $ term model suffers from some problems,  namely fine tuning as well as  coincidence problem. The fine-tuning problem indicates that if $ \Lambda $ represents vacuum energy then it is difficult to obtain the similar value of energy density of vacuum suggested by quantum field theory $ (\rho_{p})$ and $ \Lambda $ related  observed energy density of  $ (\rho_{\Lambda})$. The present ratio of energy densities $ (\rho_{\Lambda})$ and $ (\rho_{p})$ is of order $ 10^{-120}$. Similarly, the coincidence problem     
needs an explanation about effective dark energy density $ \rho_{DE} $ comparable to present normal matter density $ (\rho_{m})$~\cite{capozziello2006unified,nojiri2006oscillating,tian2020cosmological}.
In order to resolve fine-tuning and coincidence problems, several attempts have been made to  consider dynamical fields namely  quintessence~\cite{PhysRevLett.80.1582,PhysRevD.59.123504,kotambkar2014anisotropic}, phantom~\cite{caldwell2002phantom,elizalde2004late}, quintom~\cite{guo2005cosmological,zhao2006quintom,cai2010quintom} etc. for constructing a true  cosmological model representing the present-day expansion of the universe. We would also like to bring in attention to the following papers, in which the authors have explored a phantom and quintessence class of interacting model for the late Universe and showed the Universe's evolution through analysis of various density parameters~\cite{patil2023dynamics,patil2023coupled}.    
The universe's constituents are responsible for this acceleration which has been referred to as Dark Energy by researchers due to enigmatic nature. Various cosmological models involving dark energy have been proposed to elucidate this acceleration. The cosmological constant term $ \Lambda $, also studied as vacuum quantum energy, is widely considered as dark energy's promising candidate~\cite{aich2022phenomenological}. Although the $ \Lambda $ term closely aligns with observed data. It still faces two valuable challenges: first is fine-tuning  and the second is cosmological constant~\cite{copeland2006dynamics}.
\vspace{.3cm}\\
The Einstein-Hilbert action of general relativity can be extended and broadened in diverse manners. One of the simplest approaches is to incorporate a function in terms of Ricci scalar $R$ into the action, leading to the formulation of the $ f(R)$ theory, which has been  discussed in various literature~\cite{1982GReGr..14..453K, buchdahl1970non}. At later times, the cosmological expansion occurrence  can be well explained by $f(R)$ gravity~\cite{carroll2004cosmic}. For practical cosmological models, the constraints and prerequisites have been comprehensively investigated in~\cite{capozziello2006cosmological,amendola2007f}. It has been shown that there exist $f(R)$ gravity models that adhere to the constraints imposed by solar system tests~\cite{nojiri2003modified,faraoni2006solar,zhang2007behavior,amendola2008phantom}. Several authors~\cite{carroll2004cosmic,starobinsky2007disappearing,tsujikawa2008observational,capozziello2008solar,liu2018constraining} have presented observational expressions of $f(R)$ gravity models  and have also discussed the equivalence principle for  $f(R)$ gravity  and the limitations imposed  by the solar system. The astrophysical as well as cosmological implications of $f(R)$ gravity are explored in~\cite{nojiri2004gravity,allemandi2005dark,nojiri2006modified,nojiri2007newton,nojiri2007unifying,cognola2008class,nojiri2008modified,bertolami2007extra,faraoni2007viability,santos2007energy,bamba2009crossing,harko2008modified, capozziello2008cosmography,nojiri2009cosmological,faraoni2009lagrangian,nojiri2011unified,elizalde2011nonsingular,odintsov2017exponential,nojiri2017modified,nunes2017new,singh2020cosmological,mishra2021wormhole,de2023finite,odintsov2023early}. The inclusion of an explicit connection between any function of the matter Lagrangian density$(L_{m})$ and the Ricci scalar$(R)$ was introduced into the theory, extending the scope beyond that of $f(R)$ gravity theories \cite{bertolami2007extra}. This model was further developed to incorporate scenarios involving arbitrary connections in both matter and geometry \cite{harko2008modified}. The effects of non-minimal coupling between matter and geometry on cosmology and astrophysics have been extensively studied in~\cite{faraoni2007viability,faraoni2009lagrangian,nesseris2009matter,harko2010galactic,harko2010matter}.
\vspace{.3cm}\\
The $f(R, L_{m})$ gravity theory presented by Harko and Lobo~\cite{harko2010f} is basically an interesting modified theory of gravity with an arbitrary curvature-matter coupling to the maximal extension of all the gravitational theories constructed in Reimann space~\cite{nojiri2004gravity,allemandi2005dark,manna2023gravity}. The motion of test particles in $f(R, L_{m})$ gravity theory is non-geodesic and an extra force orthogonal to four velocity vector arises. This gravity models admit an explicit violation of the equivalence principle, which is highly constrained by solar system tests~\cite{Faraoni2004pi,zhang2007behavior,bertolami2008general,Rahaman2009solar,Matos2021wave}.  The field equations of $f(R, L_{m})$ gravity are equivalent to the field equations of the $f(R)$ model in empty space time, but differ from them as well as from GR, in the presence of matter~\cite{harko2010f,lobo2015extended}. Thus, the predictions of $f(R, L_{m})$ gravitational models could lead to some major differences, as compared to the predictions of standard GR, or other generalized gravity models, in several problems of current interest in cosmology.
\vspace{.3cm}\\
Although $\Lambda$CDM model has been considered as a best fit cosmological model, which adequately describes the observed expansion of the universe within the framework of general theory of relativity (GR). However, $\Lambda$CDM model has also been faced difficulties in explaining fine tuning and coincidence problems. Further, several forms of scalar fields have been considered to represent mysterious form of dark energy in GR. Again, there are various modifications of GR to find a suitable model of accelerated expansion of the universe and no one till date has claimed for the final best model of the universe. In this context, the present cosmological model is an attempt to find a suitable model in $ f (R, L_{m}) $ gravity theory. Essentially, unlike $\Lambda$CDM, this theory modifies GR to address the late time cosmic acceleration without considering the erstwhile cosmological constant as introduced by Einstein for his cosmological model to be static. To do so in a novel way it  introduces a few additional forces that deviate from the geodesic path~\cite{ShuklaDynamical}. However, in connection to observational evidences $\Lambda$CDM has a strong observational basis with some unresolved tensions whereas in this regard $ f (R, L_{m}) $ offers a strong theoretical background formalism though the theory needs observational validation in favour of its proposal.
\vspace{.3cm}\\
Based on the above motivational works, in the present investigation we have studied some cosmological issues in the framework of $ f (R, L_{m}) $ cosmological  model under data analytic techniques. This work is divided into 7 sections, which are as follows: In Sec. \ref{sec:2}, we provide a brief review in $ f(R, L_{m})$ gravity. We obtain the field equation in the FLRW metric in Sec. \ref{sec:3} whereas in the Sec. \ref{sec:4}, we begin by considering a particular  form of  $ f (R, L_{m}) $ cosmological  model. In Sec. \ref{sec:5}, we use Bayesian statistical techniques with Cosmic chronometer(CC) and Joint (CC + Pantheon) datasets to restrict the best fit values of the model parameters. In Sec. \ref{sec:6}, Deceleration parameter's graphical representation are depicted for best-fit values of parameters. We extensively discuss physical parameters like (Equation of State $(\omega)$, energy density$ (\rho) $ and pressure$\mathit{(p)}$) for best fit parameter values in this model graphically. We discuss all energy conditions and also analyze the cosmographic parameters's evolution. The Om diagnostic is investigated to distinguish our cosmological model from  dark energy's other models and discuss present universe's age. At last, in Sec. \ref{sec:7}, a summary of this research work is presented.

\section{Brief review of $ f(R, L_{m})$ gravity}\label{sec:2}

In the following section, we will provide a concise overview of the $ f(R, L_{m})$ gravity. The action in $ f(R, L_{m})$ gravity  takes the form~\cite{harko2010f}
\begin{equation}{\label{1}}
S=\int f(R, L_{m})\sqrt{-g} d^{4}x,
\end{equation}
where $\sqrt{-g}$ is the square root of the determinant of the metric tensor. $ L_{m} $ signifies the matter Lagrangian and $ R $ indicates  Ricci scalar. 
\vspace{.2cm}\\
In the context of the metric tensor ($g^{ab}$) and the Ricci tensor ($R_{ab}$), the Ricci-scalar $(R)$ is described
as shown below
\begin{equation}{\label{2}}
R=g^{ab}R_{ab}, 
\end{equation}
where the $R_{ab}$ is written in the following expression as
\begin{equation}{\label{3}}
R_{ab}= \partial_{c} \Gamma^{c}_{ab}-\partial_{a} \Gamma^{c}_{cb}+\Gamma^{c}_{ab}\Gamma^{d}_{dc}-\Gamma^{c}_{bd} \Gamma^{d}_{ac},
\end{equation}
where $ \Gamma^{\alpha}_{\beta \gamma} $ represents the Levi-Civita connection's components, which is described by
\begin{equation}{\label{4}}
\Gamma^{\alpha}_{\beta \gamma}= \frac{1}{2}g^{\alpha c}\left(\frac{\partial g_{\gamma c}}{\partial x^{\beta}}+\frac{\partial g_{c \beta}}{\partial x^{\gamma}}-\frac{\partial g_{\beta \gamma }}{\partial x^{c}} \right).
\end{equation}
On the variation of action (\ref{1}) over metric tensor $ g_{ab} $, the field equations
\begin{equation}{\label{5}}
\frac{\partial f}{\partial R}R_{ab}+(g_{ab} \square -\nabla_{a}\nabla_{b})\frac{\partial f }{\partial R}-\frac{1}{2}\left( f-\frac{\partial f}{\partial L_{m}}L_{m}\right)g_{ab}=\frac{1}{2}\left(\frac{\partial f}{\partial L_{m}}\right)T_{ab}, 
\end{equation}
where $T_{ab}$ denotes the energy-momentum tensor for perfect fluid, which is denoted by
\begin{equation}{\label{6}}
T_{ab}=\frac{-2}{\sqrt{-g}}\frac{\delta(\sqrt{-g}L_{m})}{\delta g^{ab}}.
\end{equation}
By contracting the field equation (\ref{5}), we establish the  relationship connecting the matter Lagrangian density$(L_{m})$, Ricci scalar$(R)$, and $T$ is the trace of the energy-momentum tensor$(T_{ab})$ as
\begin{equation}{\label{7}}
R\left(\frac{\partial f}{\partial R}\right)  +2\left(\frac{\partial f}{\partial L_{m}}L_{m}-f\right)+ 3\square \frac{\partial f}{\partial R}=\frac{1}{2}\left(\frac{\partial f}{\partial L_{m}}\right)T, 
\end{equation}
where $ \square I=\frac{1}{\sqrt{-g}}\partial_{a}(\sqrt{-g}g^{ab} \partial_{b}I)$ for any arbitrary function I.
\vspace{.2cm}\\
Applying the covariant derivative in equation (\ref{5}), we get
 \begin{equation}{\label{8}}
\nabla^{a}T_{ab}=2\nabla^{a}ln[f_{L_{m}}(R, L_{m})]\frac{\partial L_{m}}{\partial g^{ab}}.
\end{equation}

\section{Field equations in $ f(R, L_{m}) \text{ gravity } $}\label{sec:3}

To analyze the current cosmological model, we employ the flat FLRW metric~\cite{partridge2004introduction} within the context of a homogeneous and isotropic universe as
\begin{equation}{\label{9}}  
ds^{2}=a^{2}(t) \left( dx^{2}+ dy^{2}+ dz^{2}\right)-dt^{2},
\end{equation}
where the scale factor '$a(t)$' quantifies cosmic expansion occurring at a specific time '$t$'. 
\vspace{.2cm}\\
With regard to the metric (\ref{9}), the Christoffel symbols exhibit non-zero components as follows:
\begin{equation}{\label{10}}
\Gamma^{0}_{pq}= -\frac{1}{2}g^{00} \  \frac{\partial g_{pq}}{\partial x^{0}}, \  \  \ \ \Gamma^{r}_{0q}=\Gamma^{r}_{q0}= \frac{1}{2}g^{r\lambda} \  \frac{\partial g_{q \lambda}}{\partial x^{0}},
\end{equation}
where $ p, q, r = 1, \ 2,\ 3$.
\vspace{.2cm}\\
By utilizing equation (\ref{3}), we obtain the Ricci tensor's components that are non-zero, which are as follows
\begin{equation}{\label{11}}
R^{0}_{0}=3\frac{\ddot{a}}{a}, \  \ R^{1}_{1}=R^{2}_{2}=R^{3}_{3}=\frac{\ddot{a}}{a}+2\left(\frac{\dot{a}}{a}\right)^{2}.            
\end{equation}
Thus, the Ricci scalar obtained for the given line element (\ref{9}) is
\begin{equation}{\label{12}}
R=6 \ \left(\frac{\dot{a}}{a}\right)^{2}+6\left( \frac{\ddot{a}}{a}\right) =12H^{2} +6\dot{H},           
\end{equation}
where $ H=\frac{\dot{a}}{a} $ is the Hubble parameter.
\vspace{.2cm}\\
We consider the stress-energy momentum tensor for the perfect-fluid matter that fills the universe, corresponding to the line element (\ref{9}), as
\begin{equation}{\label{13}}
T_{ab}=(\mathit{p}+\rho )u_{a}u_{b}+pg_{ab},
\end{equation}
where  $\mathit{p}$  represents the pressure of the cosmic fluid, $ \rho $ represents energy density  and the components $ u^{a} = (1, 0, 0, 0) $ represent the four-velocity components of the cosmic perfect fluid such that $ u_{a}u^{a}=-1$. 
\vspace{.2cm}\\
The modified Friedmann equations, which elucidate the dynamics of the universe within the $f(R, L_{m})$ gravity framework can be expressed as:
\begin{equation}{\label{14}}
R^{0}_{0}\frac{\partial f}{\partial R}+\frac{1}{2}(\frac{\partial f}{\partial L_{m}}L_{m}-f)+3H\frac{\partial \dot{f}}{\partial R}=\frac{1}{2}\frac{\partial f}{\partial L_{m}}T^{0}_{0}
\end{equation}
and 
\begin{equation}{\label{15}}
R^{1}_{1}\frac{\partial f}{\partial R}+\frac{1}{2}(\frac{\partial f}{\partial L_{m}}L_{m}-f)- \frac{\partial \ddot{f}}{\partial R} -3H\frac{\partial \dot{f}}{\partial R}=\frac{1}{2}\frac{\partial f}{\partial L_{m}}T^{1}_{1}.
\end{equation}

\section{Cosmological model under $ f(R, L_{m})$ gravity theory}\label{sec:4}
We have taken a specific form of  the $ f(R, L_{m})$ gravity~\cite{harko2014generalized} for our analysis, as follows:
\begin{equation}{\label{16}}
f(R, L_{m})=\frac{R}{2}+ L_{m}^{n},
\end{equation}
where $ n $ is an arbitrary constant.
\vspace{.2cm}\\
In the context of this specific form of the $f(R, L_{m})$ gravity model with $L_{m}=\rho$~\cite{harko2015gravitational}, the Friedmann equations 
(\ref{14}) and (\ref{15}) become
\begin{equation}{\label{17}}
3H^{2}=\rho ^{n}(2n-1),
\end{equation}
and 
\begin{equation}{\label{18}}
 2\dot{H}+3H^{2}=\rho ^{n}(n-1)-n\rho ^{n-1}p.
\end{equation}
It is notable that particularly the standard Friedmann equations of general relativity can be derived for $ n=1. $
\vspace{.2cm}\\
Now, the Equation of State (EOS) parameter is define as $ \omega=\frac{p}{\rho} $, and from equation (\ref{17}) and (\ref{18}), we obtain the EoS parameter $ \omega $ as 
\begin{equation}{\label{19}}
\omega=\frac{2(2n-1)\dot{H}+3nH^{2}}{-3nH^{2}}.
\end{equation}
From the relation $ \frac{a_{0}}{a}=1+z $, we obtain $ \dot{H}=-H(1+z)\frac{dH}{dz} $ substituting this into equation (\ref{19}), we obtain 
\begin{equation}{\label{20}}
\omega=\frac{-2(2n-1)H(1+z)H'+3nH^{2}}{-3nH^{2}}.
\end{equation} 
Here we have two independent equations (\ref{17}) and (\ref{18}) while there are three unknowns namely scale factor $a(t)$, $\rho$ and $p$. In order to find an exact solution, we should have closed system of equations and hence one more physically relation among variables required. So number of independent equations and number of unknown must be same.
\vspace{.2cm}\\
An additional equation is needed in order to solve Eq. (\ref{20}) and discover the value of $H$. To address this, we utilize a parametric form for the equation of state parameter. So we have considered a specific form of  the EoS parameter for our analysis~\cite{weller2002future} 
\begin{equation}{\label{95}}
\omega (z) =w_{0}+w_{1}z,
\end{equation}
where $w_{0}$ and $w_{1}$ are model parameters.
\vspace{.2cm}\\
The parametrization employed in this study offers several key advantages in the context of modelling the EOS of Dark Energy (DE). Firstly, it provides a physically interpretable framework, with $w_{0}$ representing the Dark energy (DE) EOS at the present epoch $(z=0)$ and $w_{1}$ characterizes its dynamics. This transparency enhances our understanding of the underlying physical processes. Moreover, the parametrization explicitly incorporates the redshift dependence, allowing us to capture the evolving nature of DE over cosmic time. Its flexibility enables us to encompass a wide range of DE behaviours from quintessence to phantom DE. Therefore, it is viable choice for studying this specific EOS.
\vspace{.3cm}\\
From equation (\ref{20}) and equation (\ref{95})
\begin{equation}{\label{96}}
w_{0}+w_{1}z=\frac{-2(2n-1)H(1+z)H'+3nH^{2}}{-3nH^{2}}.
\end{equation}
Solving equation (\ref{96}), the Hubble parameter's expression is obtained in relation to red-shift as follows
\begin{equation}{\label{21}}
H(z)=h_{0}(1+z)^{\frac{3n(w_{0}+w_{1}+1)}{2(2n-1)}}exp\left(\frac{3nzw_{1}}{2(2n-1)}\right),
\end{equation}
where the Hubble parameter's current value is signified by $h_{0}$.

\section{Observational constraints}\label{sec:5}

This section examines the compliance of parameterized from the Hubble parameter with the datasets for the cosmic chronometer (CC) and the joint data composed of cosmic chronometer and Pantheon nomenclature as CC+Pantheon. In order to perform statistical analysis, we use $ \chi^{2}$ minimization and Bayesian statistical techniques with the Markov Chain Monte Carlo (MCMC) method, which is implemented in the emcee Python library~\cite{foreman2013emcee} to constrain the model parameters best fit value to study of the cosmic behaviour in the model. The cosmological studies based on parametrizations of different parameters are widely explored \cite{mandal2024late, mandal2023cosmic}. 

\subsection{The Cosmic chronometer data}\label{sec:5.1}

\begin{center}
\begin{figure}
\includegraphics[scale=0.60]{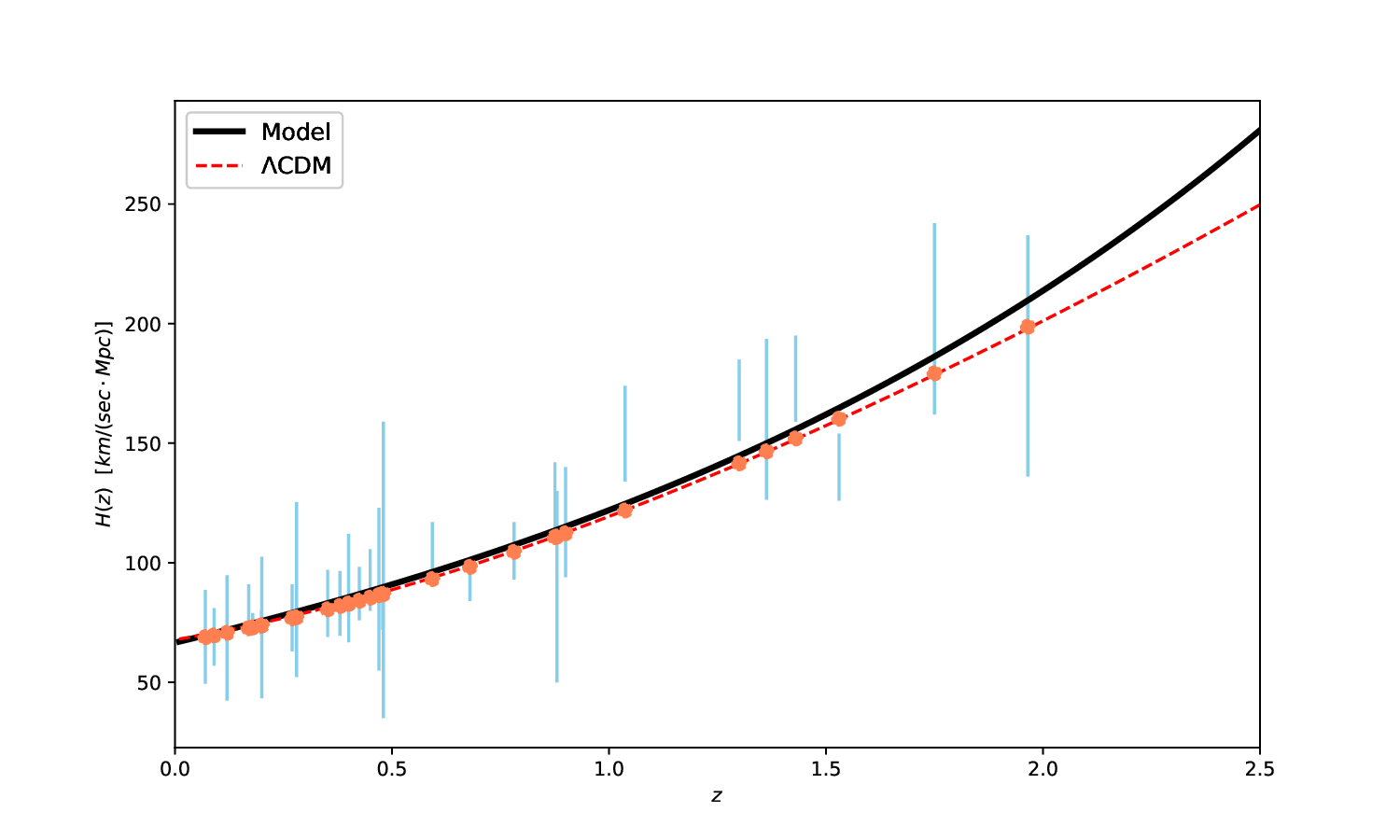}
\caption{Best fit curve of Hubble parameter $H(z)$ versus $\mathit{z} $ for the present model compared to the $\Lambda CDM$ model}
\label{fig:5}
\end{figure}
\end{center}

\begin{center}
\begin{figure}
\includegraphics[scale=0.69]{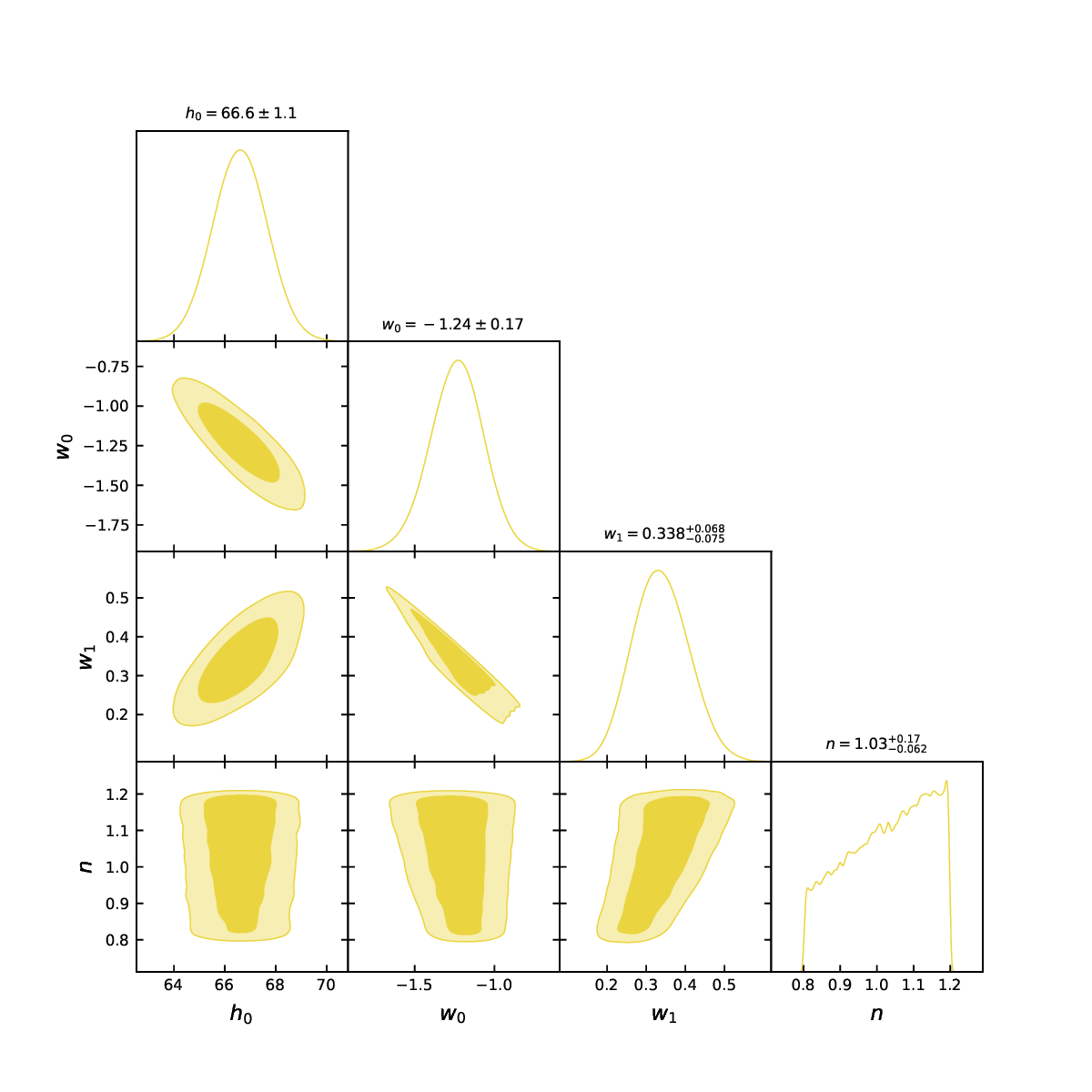}
\caption{$1\sigma$ and $2\sigma$ contour map for $h_{0}$, $w_{0}$, $w_{1}$ and $n$ using CC data set.}
\label{fig:6}
\end{figure}
\end{center}

In this part, we examine the observational implications of our cosmological model. We employ a dataset comprising 31 Cosmic Chronometer data points~\cite{simon2005constraints,sharov2018predictions} obtained through the differential ages (DA) of galaxies method within the redshift range $0.07 \leq z \leq 1.965$ to constrain the model parameters best fit value. In basic principle for the cosmic chronometer observations proposed by Jimenez and Loeb \cite{jimenez2002constraining} relates the hubble parameter$(H)$, cosmic time$(t)$ and red-shift$(z)$ as $H(z)=\frac{-1}{(1+z)}\frac{dz}{dt} $. The model parameters $h_{0}, w_{0}, w_{1}$ and $ n $ can be determined based on observational constraints by minimizing the $\chi^{2}$ function (i.e., equivalent to maximizing the likelihood function) which is expressed as
\begin{equation}{\label{27}}
\chi^{2}_{CC}(\theta)=\sum_{i=1}^{i=31} \frac{[H_{th}(\theta,z_{i})-H_{obs}(z_{i})]^{2}    }{ \sigma^{2}_{H(z_{i})}}.   
\end{equation} 
\vspace{0.1cm}\\
Here the theoretical value of Hubble is represented by $H_{th}$, the observed value of Hubble is represented by $H_{obs}$, $\sigma_{H}$ represents the observed value's standard error.
\vspace{0.1cm}\\
To compare the $H(z)$ of the current model with that of the $ \Lambda$ Cold dark matter $(\Lambda CDM)$ model, the error bars of $31$ CC data points are  illustrated in figure $(\ref{fig:5}) $ for the Hubble parameter specified in equation (\ref{21}). Figure $(\ref{fig:6})$ depicts  a contour map with $1\sigma$ and $2\sigma$ confidence levels for the constrained values of $h_{0}$, $w_{0}$, $w_{1}$ and $n$ using CC data.

\subsection{The Pantheon data}\label{sec:5.2}
\begin{center}
\begin{figure}
\includegraphics[scale=0.67]{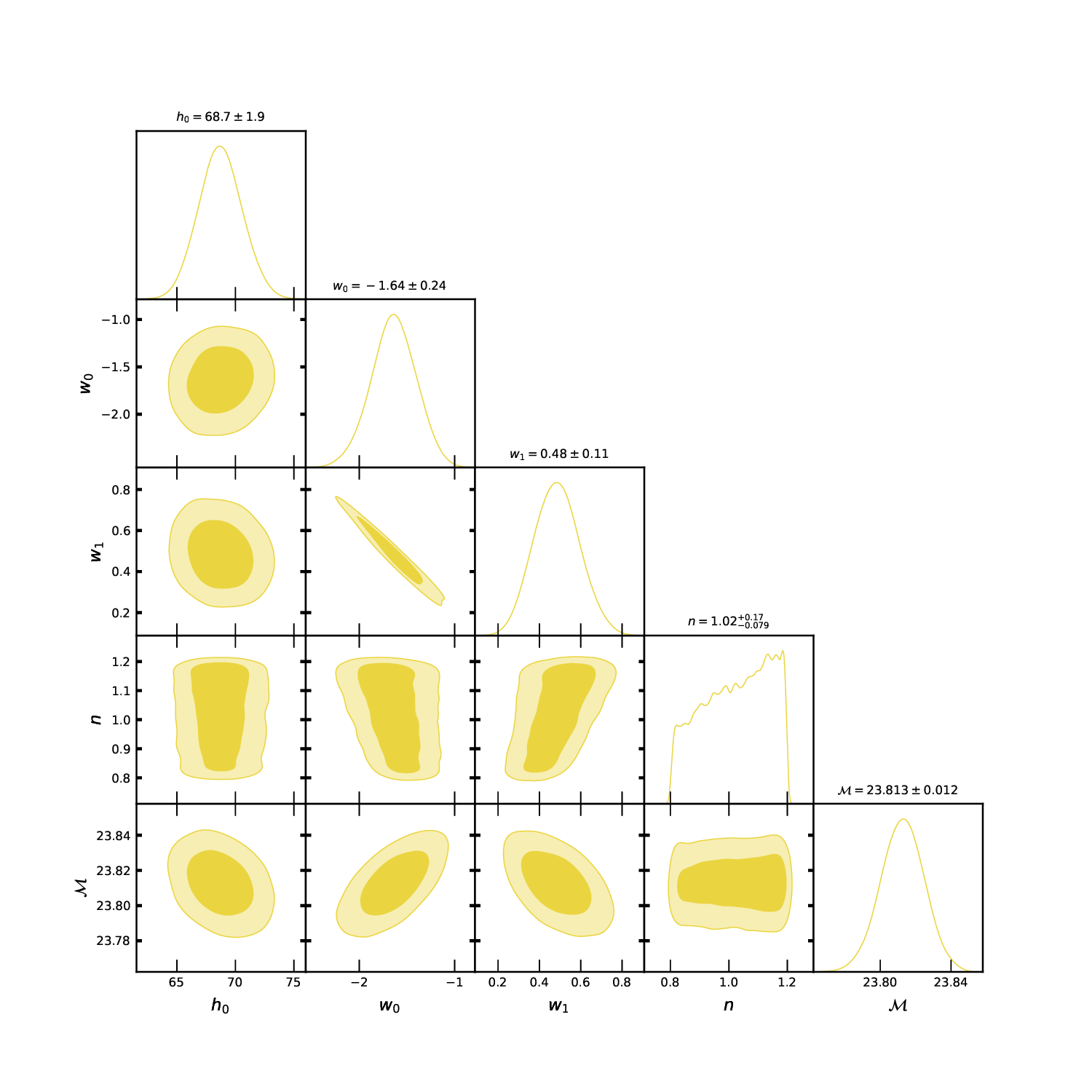}
\caption{$1\sigma$ and $2\sigma$ contour map for $h_{0}$, $w_{0}$, $w_{1}$ and $n$ using CC+Pantheon data set.}
\label{fig:7}
\end{figure}
\end{center}

The Pantheon sample, which includes $1048$ SNIa data points for the red-shift range $0.01 < z < 2.26$ is collected in Ref.~\cite{scolnic2018complete}. The CfA1-CfA4~\cite{riess1999bvri,hicken2009improved} surveys, Pan-STARRS1 Medium Deep Survey~\cite{scolnic2018complete}, SDSS~\cite{sako2018data}, SNLS~\cite{guy2010supernova}, Carnegie Supernova Project (CSP)~\cite{contreras2010carnegie} support the SN Ia sample.
\vspace{.2cm}\\
The theoretically expected apparent magnitude $\mu_{th}$(z) is given by
\begin{equation}{\label{7a}}
\mu_{th}(z)=M+5log_{10}\left[\frac{d_{L}(z)}{Mpc}\right]+25,
\end{equation}
where $M$ is the absolute magnitude. Also, the luminosity distance $d_{L}(z)$ may be defined as~\cite{odintsov2018cosmological}
\begin{equation}{\label{8a}}
d_{L}(z)=c(1+z)\int_{0}^{z}\frac{dz'}{H(z')},
\end{equation}
where $z$ represents SNIa's red-shift as  determined in the cosmic microwave background (CMB) rest frame and and $c$ is the speed of light. 
\vspace{.2cm}\\
The luminosity distance $(d_{L})$ is typically substituted with the Hubble-free luminosity distance$(D_{L}(z) \equiv H_{0}d_{L}(z)/c)$. Observe that the Hubble free luminosity distance is dimensionless, whereas the luminosity distance has a dimension called Length. The equation (\ref{7a}) could also be rewritten as
\begin{equation}{\label{9a}}
\mu_{th}(z)=M+5log_{10}\left[D_{L}(z)\right]+5log_{10}\left[\frac{c/H_{0}}{Mpc}\right]+25. 
\end{equation}
The parameters $M$ and $H_{0}$ can be combined to create a new parameter $\mathcal{M}$, which can be identified as
\begin{equation}{\label{10a}}
\mathcal{M}\equiv M+5log_{10} \left[\frac{c/H_{0}}{Mpc}\right]+25=M-5log_{10}(h)+42.38, 
\end{equation}
where $H_{0}=h \times 100$ Km/s/Mpc. We use this parameter with pertinent $\chi^{2}$ for Pantheon data in the MCMC analysis as~\cite{asvesta2022observational}
\begin{equation}{\label{11a}}
\chi^{2}_{P}= \nabla \mu_{i}C^{-1}_{ij}\nabla \mu_{j},
\end{equation}
where $\nabla \mu_{i}=\mu_{obs}(z_{i})-\mu_{th}(z_{i})$, $C_{ij}^{-1}$ is the inverse of covariance matrix and $\mu_{th}$ will be given by equation (\ref{9a}). 
\vspace{.2cm}\\
The luminosity distance depends on the Hubble parameter. Therefore, we use the emcee package~\cite{foreman2013emcee} and equation (\ref{21}) to get the maximum likelihood estimate with the joint CC+Pantheon dataset. The joint $\chi^{2}$ for maximum likelihood estimate is defined as $\chi^{2}_{CC}+\chi^{2}_{P}$. Figure $(\ref{fig:7})$ is illustrated the the marginalized $1\sigma$ and $2\sigma$ contour map and 1D posterior distribution from the Monte Chain Monte Carlo analysis using CC+Pantheon data.
\vspace{.2cm}\\
For the CC data set, we use $32$ random chains (walkers) and 40000 iterations (steps) for the MCMC analysis. We select uniform priors on $h_{0}, w_{0}, w_{1}$ and $n$ for the CC data set. We take the range $40 < h_{0} < 90, -3 < w_{0} <0, 0 < w_{1} <1 $ and $  0.8 < n < 1.2$. For the joint data set, we use $48$ random chains (walkers) and 10000 iterations (steps) for the MCMC analysis. We select uniform priors on $h_{0}, w_{0}, w_{1}, n$ and $\mathcal{M}$ for the joint data set. We take the range $40 < h_{0} < 90, -3 < w_{0} <0, 0 < w_{1} <1, 0.8 < n < 1.2$ and $ 23.5 < \mathcal{M} < 23.95 $
\vspace{.2cm}\\
Table (\ref{table:1}), summarizes the best fit values for the model parameters.

\begin{table}[htbp]
\centering
\begin{tabular}{|c|c|c|c|c|c|}
\hline
Dataset & $h_{0}$[Km/(s.Mpc)] & $w_{0}$ & $w_{1}$ & $n$ & $\mathcal{M}$ \\
\hline
CC & $66.6^{+1.1}_{-1.1}$ & $-1.24^{+0.17}_{-0.17}$ & $0.338^{+0.068}_{-0.075}$ &$1.03^{+0.17}_{-0.062}$ & - \\
\hline
CC+Pantheon  & $68.7^{+1.9}_{-1.9}$  & $-1.64^{+0.24}_{-0.24}$  & $0.48^{+0.11}_{-0.11}$ &  $1.02^{+0.17}_{-0.079}$ & $23.813^{+0.012}_{-0.012}$\\
\hline
\end{tabular}
\caption{Model parameters's best fit values in MCMC analysis}
\label{table:1}
\end{table}

\section{Physical and dynamical properties of the model}\label{sec:6}

\subsection{Deceleration parameter}\label{sec:6.1}

In defining how the cosmos is expanding, the deceleration parameter $(q)$ plays a crucial role. It depicts the behavior of the cosmic universe. Based on the values of the deceleration parameter$(q)$, the universe's expansion history can be characterized: $ q < 0 $ for the accelerated phase and $ q > 0 $ for the decelerated phase. The universe undergoes a phase of super-accelerated expansion when the deceleration parameter is less than  $-1 $. The values of $ q $, which are $ -1, \frac{1}{2}, $ and $ 1 $, correspond to different phases of the universe: de Sitter, matter-dominated, and radiation-dominated phases, respectively.  In other words, the rate of universe expansion can be described using the deceleration parameter specified by $ q=\frac{-\ddot{a}}{aH^{2}} $. Alternatively, the relation can also be expressed as
\begin{equation}{\label{22}}
 q = -1 + \frac{d}{dt}\frac{1}{H}.
\end{equation}
Substituting  (\ref{21}) into (\ref{22}), we obtain
\begin{equation}{\label{23}}
q(z)=-1+{\frac{3n[w_{0}+w_{1}(2+z)+1]}{2(2n-1)}}.
\end{equation}
\begin{figure}
\includegraphics[width=15.5cm,height=6.5cm]{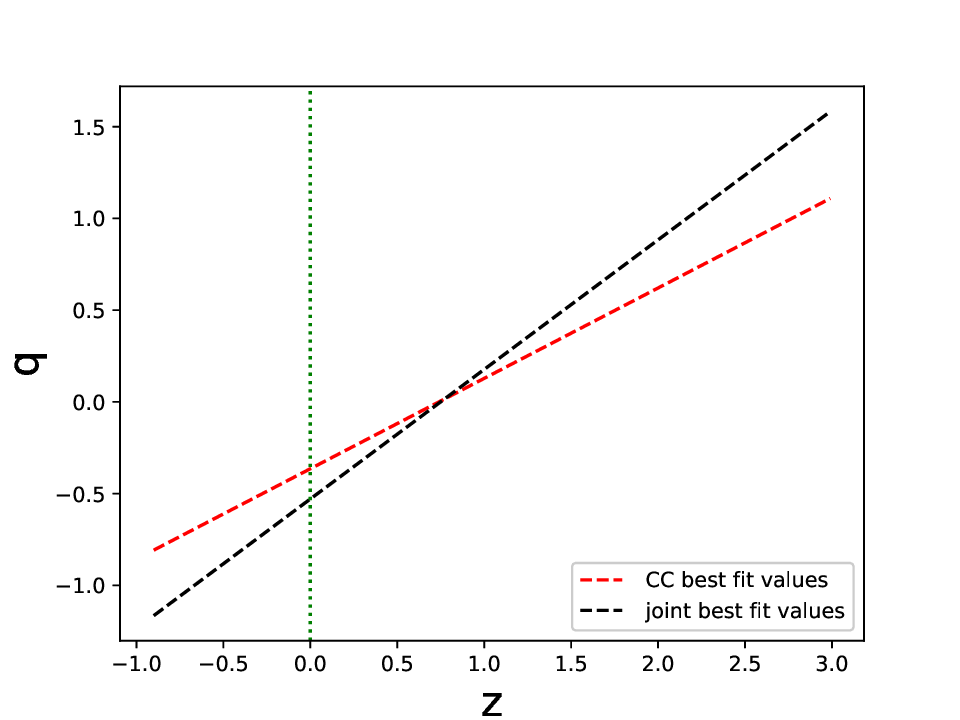}
\caption{deceleration parameter $(\mathit{q})$ with $\mathit{z}$ for constrained model parameters using CC data set and joint data set}
\label{fig:1}
\end{figure}
For CC data set and joint data set, the analysis of deceleration parameters as shown in $ (\ref{fig:1})$. Figure $ (\ref{fig:1}) $ illustrates the transition of the universe from a decelerated expansion phase to an accelerated expansion phase at $ z= 0.747 $( for $w_{0}=-1.24, w_{1}=0.338, n=1.03$ best fit values from CC data set) and $ z= 0.72$(for $w_{0}=-1.64, w_{1}=0.48, n=1.02$ best fit values from joint data set) which is in good agreement with  $ 0.60 \leq z_{t}\leq 1.18 $ ($ 2\sigma $, joint analysis)~\cite{lima2012transition} and the present values of deceleration parameters are $q_{0}=-0.36$~\cite{yadav2024reconstructing}( for $w_{0}=-1.24, w_{1}=0.338, n=1.03$ best fit values form CC data set) and $q_{0}=-0.50$ (for $w_{0}=-1.64, w_{1}=0.48, n=1.02$ form joint data set). This negative value indicates that the expansion of the universe is accelerating at this time.

\subsection{Energy Density and Pressure}\label{sec:6.2}
\begin{figure}[!htb]
\captionsetup{skip=0.4\baselineskip,size=footnotesize}
   \begin{minipage}{0.40\textwidth}
     \centering
     \includegraphics[width=8.5cm,height=5.5cm]{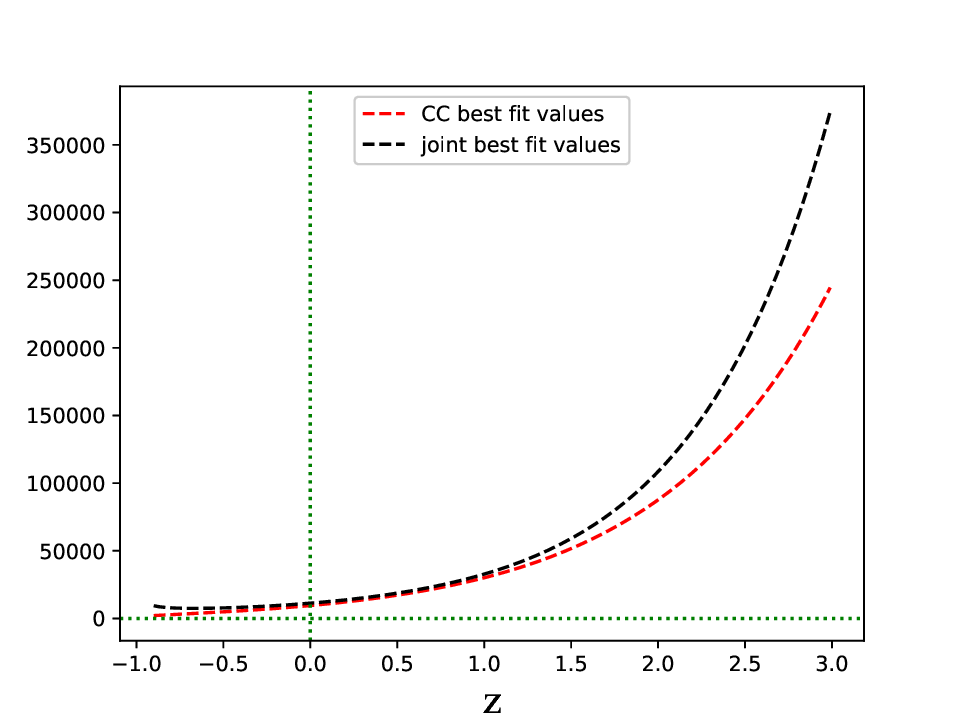}
\caption{Energy density $(\rho)$ with $\mathit{z}$ for constrained model parameters using CC data set and joint data set}
\label{fig:2}
    \end{minipage}\hfill
   \begin{minipage}{0.40\textwidth}
     \centering
     \includegraphics[width=8.5cm,height=5.5cm]{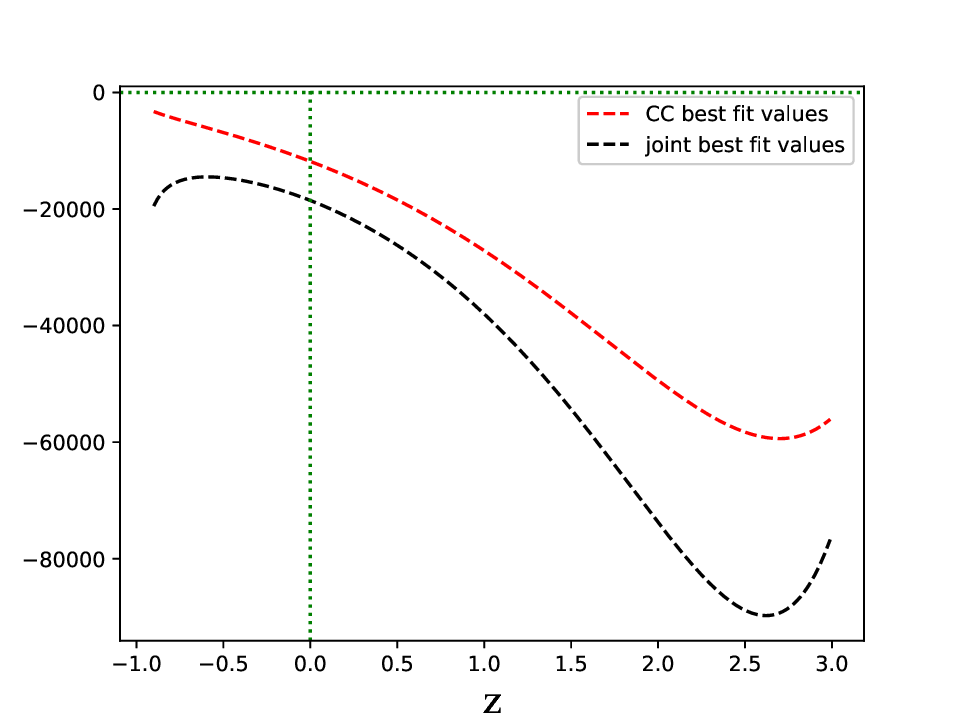}
  \caption{pressure $(\mathit{p})$ with $\mathit{z}$ for constrained model parameters using CC data set and joint data set}
\label{fig:3}
   \end{minipage}
\end{figure}

In the expansion history, the energy density remains positive, while the pressure may become negative from the recent past. The  pressure being negative from the recent past may yield
the mechanism for accelerating universe expansion in model. 
The expansion history during the decelerating era is depicted by a positive, decreasing energy density, As the transition from deceleration to acceleration occurs, the energy density continues to stay positive, while the pressure becomes negative due to the dominance of dark energy.
\vspace{.2cm}\\
From equation (\ref{17}) and (\ref{21}), we obtain the energy density 
\begin{equation}{\label{24}}
\rho (z)= \left(\frac{3h_{0}^{2}(1+z)^{\frac{3n(1+w_{0}+w_{1})}{2n-1}}exp\left(\frac{3nzw_{1}}{2n-1}\right)}{2n-1}\right)^{\frac{1}{n}}.  
\end{equation}
Using Eqs. (\ref{18}),~(\ref{21}) and (\ref{24}), the cosmic pressure is obtained as  
\begin{equation}{\label{25}}
\mathit{p}(z)=[w_{0}+w_{1}.z]\left(\frac{3h_{0}^{2}(1+z)^{\frac{3n(1+w_{0}+w_{1})}{2n-1}}exp\left( \frac{3nzw_{1}}{2n-1}\right)}{2n-1}\right)^{\frac{1}{n}}.
\end{equation}
For constrained model parameters $h_{0}=66.6, w_{0}=-1.24, w_{1}=0.338, n=1.03$ (best fit values from CC data set) and $h_{0}=68.7, w_{0}=-1.64, w_{1}=0.48, n=1.02$ (best fit values from joint data set), the evolution of energy density and pressure has been illustrated in Figure $(\ref{fig:2})$ and $(\ref{fig:3})$ respectively. It is noteworthy that the energy density exhibits the expected positive behavior, indicating a contribution to the Universe’s expansion, while the pressure displays a negative behavior both in the present and future. These observations align with the expanding nature of the accelerating Universe.
 \vspace{0.3cm}\\
It has been observed that the energy conservation equation $\dot{\rho}+3H(\rho+p)=0$ is satisfied for obtained equation of $H(z)$, $\rho(z)$ and $p(z)$ of our model that is stated in equation (\ref{21}), (\ref{24}) and (\ref{25}) respectively. This indicates that total energy is conserved for this model.

\subsection{Equation of State Parameter}\label{sec:6.3}

The relationship between pressure $(\mathit{p})$ and energy density $(\rho)$ within the universe is described by the EOS parameter. The dust phase $(\omega=0)$, the radiation dominated phase $(\omega=\frac{1}{3}) $ and the vacuum energy phase $(\omega=-1)$, corresponding to the $\Lambda$CDM model are some of the common phases observed through the EOS parameter. Furthermore, there is the Universe's accelerating phase, which has been the subject of recent discussions and is denoted by $ (\omega < \frac{-1}{3})$. The quintessence regime $ (-1 < \omega < \frac{-1}{3})$ and the phantom regime $(\omega < -1)$ are included in this phase.
\vspace{0.2cm}\\
The EoS parameter $ (\omega=\frac{\mathit{p}}{\rho}) $ can be expressed using (\ref{24}) and (\ref{25}) as:
\begin{equation}{\label{26}}
\omega=w_{0}+w_{1}z.
\end{equation}
\begin{center}
\begin{figure}
\includegraphics[width=15.5cm,height=5.5cm]{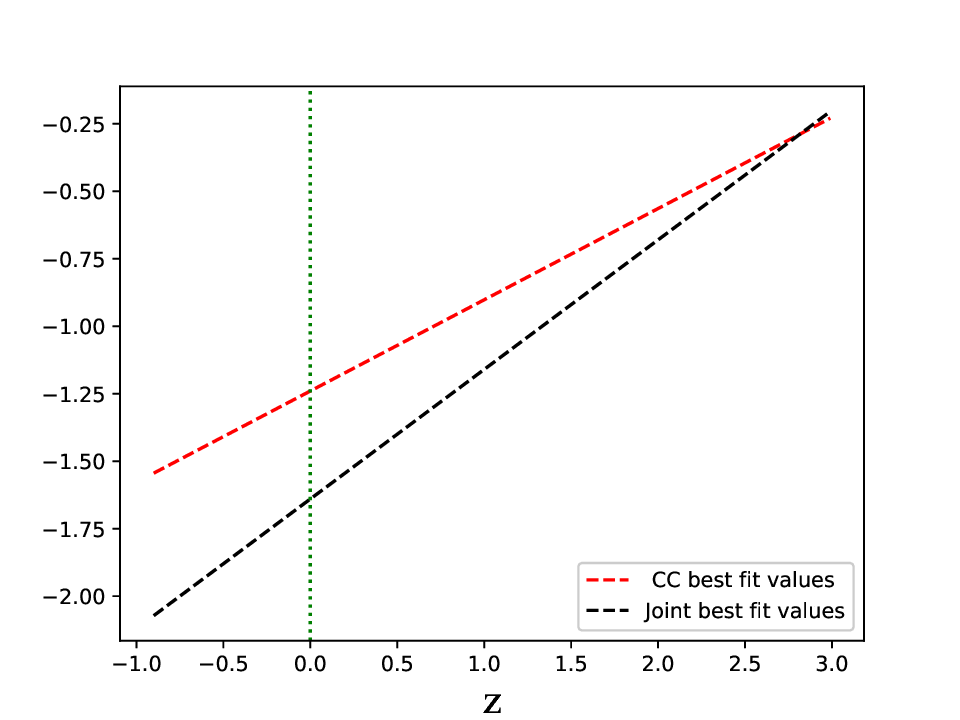}
\caption{Equation of State parameter $ (\omega) $  with $\mathit{z}$ for constrained model parameters using CC data set and joint data set}
\label{fig:4}
\end{figure}
\end{center}
The EOS parameter (which is given in equation (\ref{26})) is examined in this study. Figure $(\ref{fig:4})$ illustrates the behaviour of the EOS parameter (which is given in equation (\ref{26})) based on the constrained values of these parameters $w_{0}$ and $w_{1}$ from the CC data set and joint data set.  At $z=0$, EOS parameter's values are found to be
$\omega=-1.24$ (for $w_{0}=-1.24, w_{1}=0.338$ best fit values from CC data set) and $\omega=-1.64$  (for $w_{0}=-1.64, w_{1}=0.48$ best fit values from joint data set) respectively.

\subsection{Energy conditions}\label{sec:6.1}

\begin{figure}[!htb]
\captionsetup{skip=0.4\baselineskip,size=footnotesize}
   \begin{minipage}{0.40\textwidth}
     \centering
     \includegraphics[width=8.cm,height=7cm]{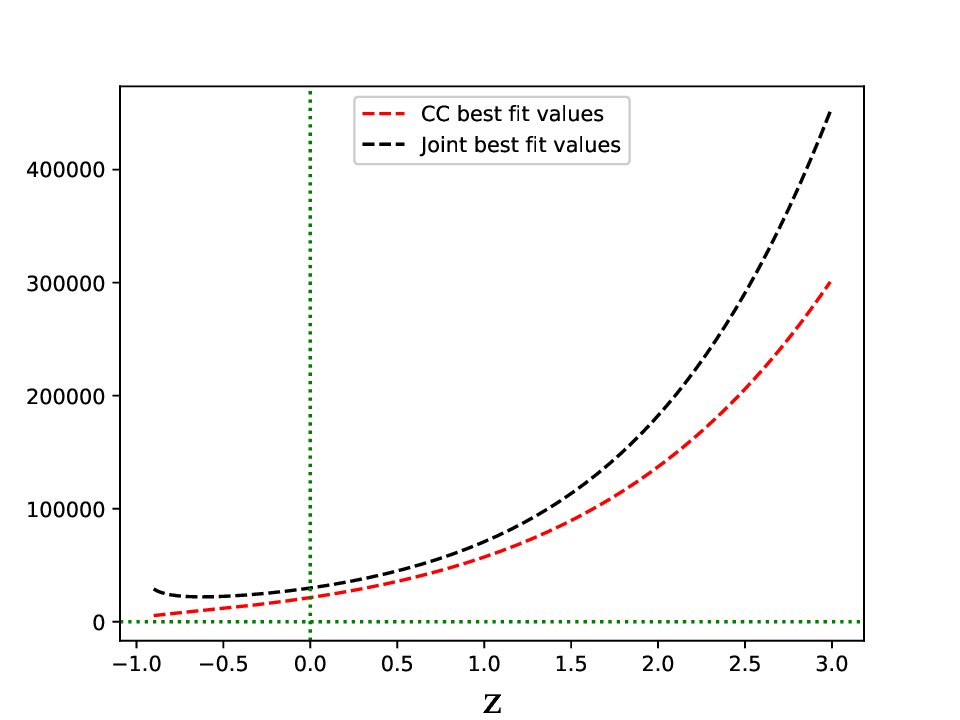}
\caption{Profile of $(\rho - p)$ with $\mathit{z}$ for constrained model parameters using CC data set and joint data set}
\label{fig:8}
    \end{minipage}\hfill
   \begin{minipage}{0.40\textwidth}
     \centering
     \includegraphics[width=8.cm,height=7cm]{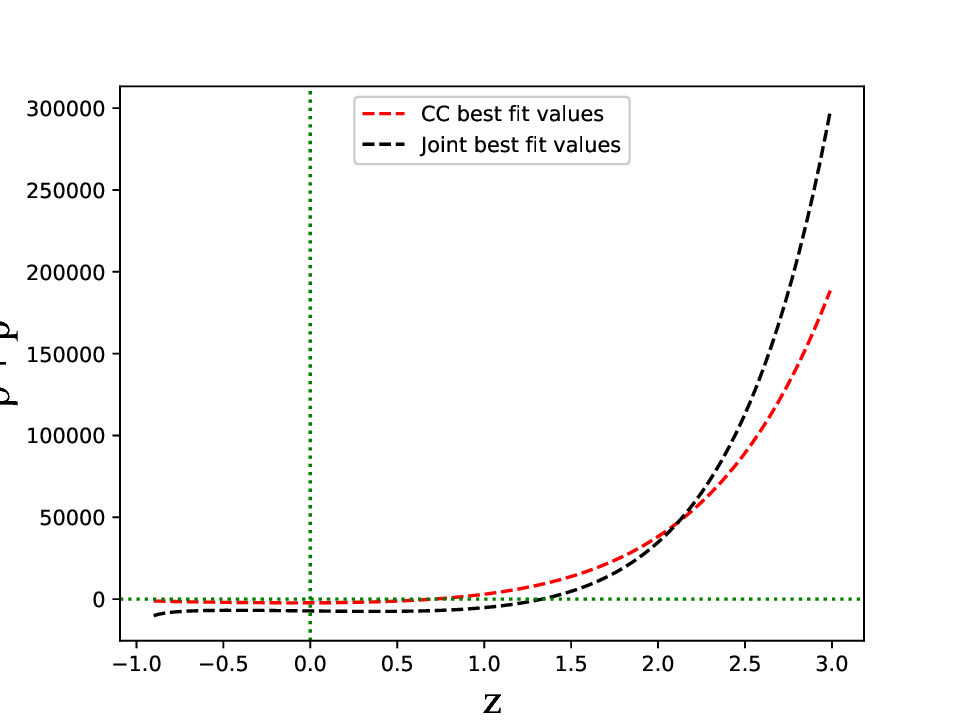}
  \caption{Profile of $(\rho + p)$ with $\mathit{z}$ for constrained model parameters using CC data set and joint data set }
\label{fig:9}
   \end{minipage}
\end{figure}

\begin{figure}
\includegraphics[width=11.5cm,height=4.5cm]{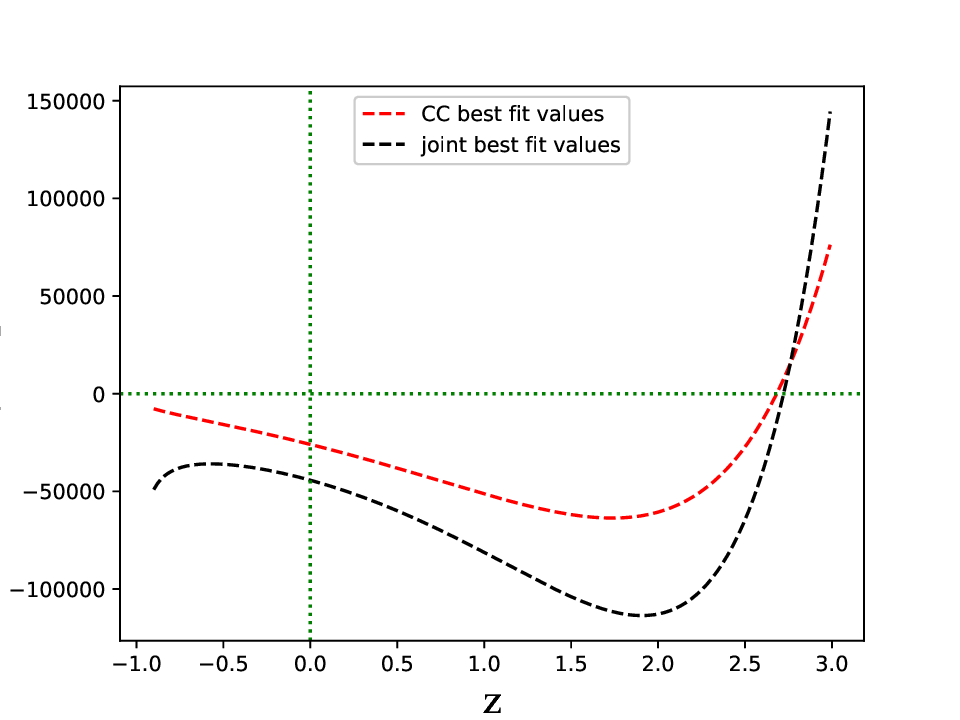}
\caption{Profile of $(\rho+3p)$ with $\mathit{z}$ for constrained model parameters using CC data set and joint data set }
\label{fig:10}
\end{figure}

Energy conditions are the coordinate invariant restriction on the stress-energy tensor of the model.  The Raychaudhuri equation serves as the foundational source for the strong energy condition and may assert for gravity's attractive nature ~\cite{singh2022lagrangian}. These conditions explore the coherence of null, light-like, space-like and time-like geodesics, playing a crucial role in characterizing the fluid that is dealing~\cite{singh2022cosmological, bouhmadi2015little}. The important conditions on energy density and pressure of the universe in a model are as follows~\cite{singh2022lagrangian}:
\vspace{.3cm}\\
$(1)$ The sum of energy density and pressure is always non-negative as measured by null like observer $(\rho + p \geq 0) $. This condition is known as Null energy condition.
\vspace{.2cm}\\
$(2)$ The weak energy condition states that the energy density as well as the sum of energy density and pressure are non-negative $(\rho \geq 0, \  \ \rho + p \geq 0) $
\vspace{.2cm}\\
$(3)$ According to the dominant energy condition, the energy density is always non-negative and $( \rho  \geq |p| ) $.
\vspace{.2cm}\\
$(4)$ The strong energy condition states that the active gravitational mass $(\rho + 3p)$ is positive and $(\rho + p)$ must be non-negative.
\vspace{.3cm}\\
During the decelerating phase, the active gravitational mass $ (\rho + 3p) $ remains positive. however, observational data suggests a violation of this condition sometime between the epoch of galaxy formation and the present. Hence, the expansion of the universe at an accelerating rate (and consequently the presence of repulsive gravity) may be experienced for $ \rho + 3p < 0 $. The inequality $ \rho + 3p < 0 $ indicates the potential existence of negative cosmic pressure, which exhibits anti-gravitational characteristics. It's important to mention that the SEC comprises two inequalities, so the violation of either one would result in a violation of the SEC \cite{singh2022lagrangian,singh2023homogeneous,lalke2023late}.
Figures $ (\ref{fig:8}), (\ref{fig:9})$ and $(\ref{fig:10}) $ depict the graphical depiction of energy conditions for the considered  model $f(R, L_{m})=\frac{R}{2}+L_{m}^{n}$. 
\vspace{.2cm}\\
In both present and future, $(\rho+p)$ and  $(\rho+3p)$ attain negative values while $(\rho-p)$ remains positive. The violation of NEC indicates the presence of exotic matter. For ordinary matters, NEC is satisfied. In classical general relativity, wormholes are supported by exotic matter, which involves a stress-energy tensor that violates the null energy condition (NEC)\cite{morris1988wormholes,visser2000energy}. 
\vspace{0.3cm}\\
Furthermore, the trajectory of $ (\rho + 3p) $  shifts  from positive to negative values around $z=2.68$ {\bf{(for CC data set)}} and $z=2.72$ {\bf{(for joint data set)}}.
\vspace{.2cm}\\
The validity of null energy condition points for the existence of either the decelerating or the accelerating expansion due to the quintessence kind of dark energy. For the constrained values according to CC data sets the $(\rho+P) \geq 0$. However, for the constrained values according to jonit data, the violation of $(\rho+P) \geq 0$ points for the presence of phantom kind of dark energy in the model. In summory, we may conclude that the phantom kind of dark energy may be ruled out according to violation of $(\rho+P) \geq 0$ in the model.

\newpage

\subsection{Cosmographic parameters}\label{sec:6.2}

This section is dedicated to examining cosmography's role in cosmology. Its objective is to examine kinematic quantities that characterize the cosmological scenario. Due to this, it is alternatively referred to as cosmo-kinetics or the kinematics of the Universe. Weinberg was the first to discuss the cosmography paradigm~\cite{weinberg2008cosmology}, and he introduced the scale factor that increased around current time $ t_{0} $ by utilizing a Taylor series.
\vspace{.2cm}\\
Before obtaining evidence of the universe's accelerated expansion, cosmologists considered $\mathit{H}$ as a variable observational quantity. Indeed, $\mathit{q}$   illustrates its progression, which is the second-order derivative of `$a$' \cite{mukherjee2016parametric}. The jerk $\mathit{(j)}$ and snap $\mathit{(s)}$ parameters offer insights into the cosmic development of the cosmos. Jerk is occasionally also called `jolt', while `jounce' is another name of  snap~\cite{visser2004jerk}. They are defined in~\cite{visser2004jerk, visser2005cosmography}  as follows:
\begin{equation}{\label{28}}
\mathit{j}=\frac{1}{aH^{3}}\left(\frac{d^{3}a}{dt^{3}}\right),\ \ \mathit{s}=\frac{1}{aH^{4}}\left(\frac{d^{4}a}{dt^{4}}\right)
\end{equation}
\vspace{.1cm}\\
The above equation (\ref{28}) can be reformulated for $\mathit{j}$  and $\mathit{s}$  in terms of redshift~\cite{wang2009probing} as
\begin{equation}{\label{29}}
j(z)=\frac{dq}{dz}(1+z)+q(z)\left(1+2q(z)\right), \  \ \  \   s(z)=-\frac{dj}{dz}(1+z)-j(z)\left(2+3q(z)\right).
\end{equation}

\begin{figure}[!htb]
\captionsetup{skip=0.4\baselineskip,size=footnotesize}
   \begin{minipage}{0.40\textwidth}
     \centering
     \includegraphics[width=8.cm,height=7cm]{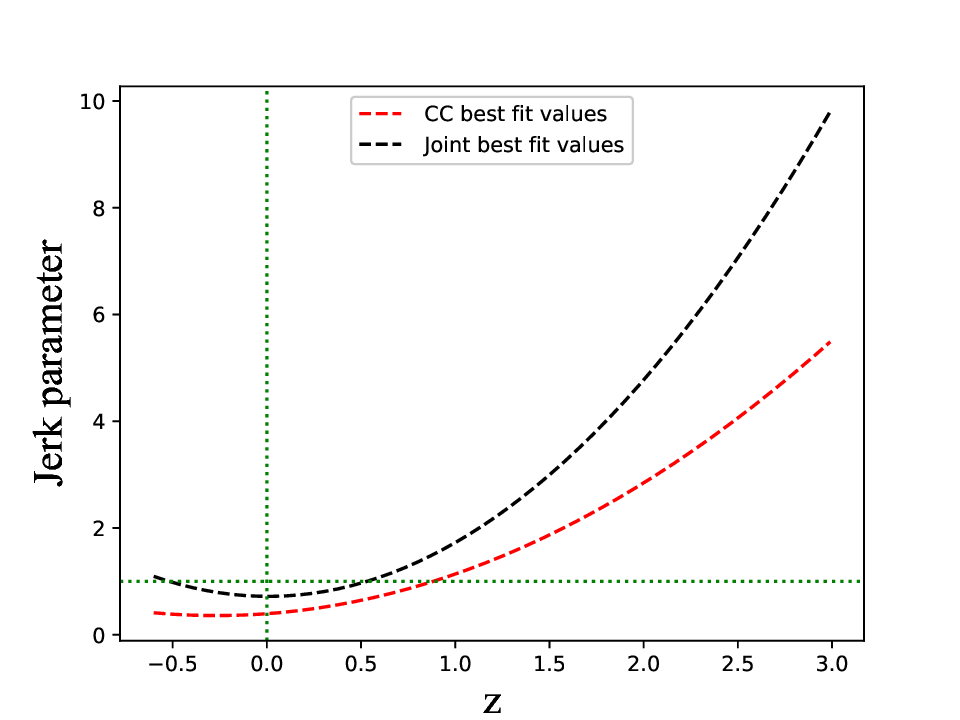}
\caption{Profile of jerk parameter $\mathit{j}$ with $\mathit{z}$ for constrained model parameters using CC data set and joint data set}
\label{fig:11}
    \end{minipage}\hfill
   \begin{minipage}{0.40\textwidth}
     \centering
     \includegraphics[width=8.cm,height=7cm]{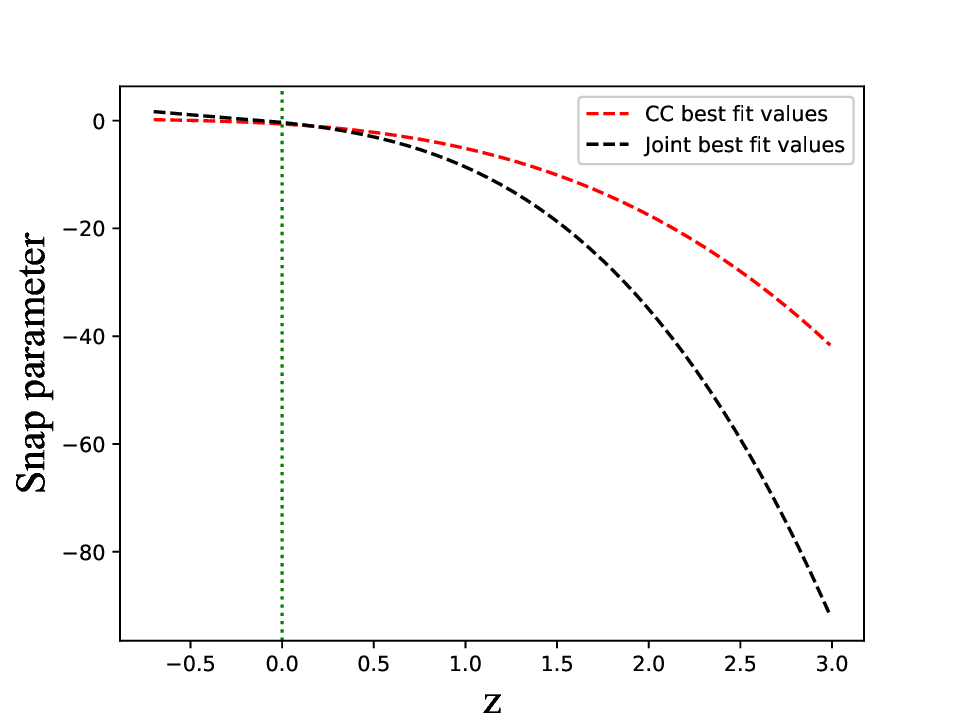}
  \caption{Profile of snap parameter $\mathit{s}$ with $\mathit{z}$ for constrained model parameters using CC data set and joint data set }
\label{fig:12}
   \end{minipage}
\end{figure}
The jerk parameter represents the rate of change of universe acceleration (or deceleration) over the time.  The sign of the jerk parameter decides the changes in the dynamics of the universe. The positive value of the jerk parameter denote the occurrence of the transitional intervals when the universe's expansion changes.
The jerk parameter's value is positive $(j > 0)$ as can be seen in Figure $ (\ref{fig:11})$. It indicates the existence of a transitional period when the universe changes its phase of expansion. The snap parameter measures the rate at which jerk parameter is changing. The behaviour of the jerk and snap parameters  are depicted in Figures $ (\ref{fig:11})$ and  $(\ref{fig:12})$. Figure $ (\ref{fig:11})$ illustrated that from early times to late time the values of jerk parameter decreases  for best fit values of the model parameters. At late time evaluation of the universe this model is similar to $\Lambda$CDM model for both datasets, viz. CC data and CC+Pantheon data.
\vspace{.2cm}\\
The behavior of the snap parameters is depicted in Figure  $(\ref{fig:12})$. It is evident that in the early universe, the snap parameter ($\mathit{s}$) evolves within the negative range and undergoes a transition from negative values to positive values as time progresses from higher to lower red-shift.

\subsection{Om diagnostics}\label{sec:6.3}

Om diagnostic is useful equipment for categorizing the dark energy's cosmological models \cite{sahni2008two}. For a flat universe, it is defined as
\begin{equation}{\label{30}}
Om(z)=\frac{\frac{H^{2}(z)}{h_{0}^{2}}-1}{(1+z)^{3}-1}.
\end{equation}
\vspace{.2cm}\\
Om(z)'s positive slope implies the phantom-like behavior of dark energy, whereas the negative slope indicates the quintessence-like behavior of dark energy. The constant behavior of Om(z) signifies the $ \Lambda$CDM model.
\begin{figure}
\includegraphics[width=11.5cm, height=6.5cm]{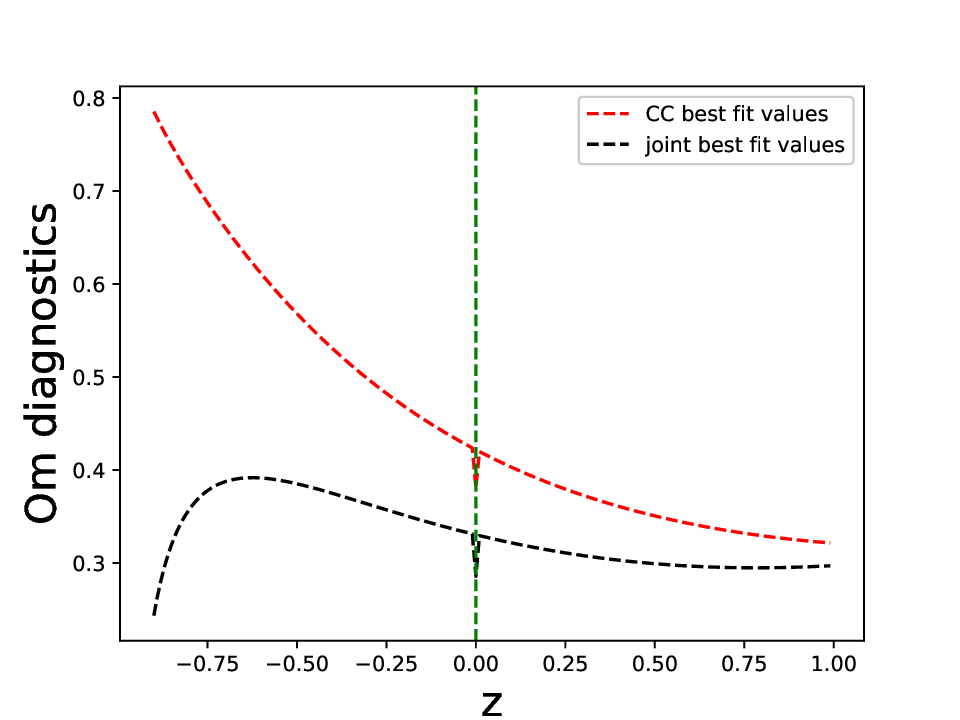}
\caption{Om diagnostics with $\mathit{z}$ for constrained model parameters using CC data set and joint data set}
\label{fig:13}
\end{figure}
The behavior of the Om(z) curve is depicted in figure $(\ref{fig:13})$. Hence, It is clear that from Figure $(\ref{fig:13})$ that this $ f(R, L_m) $ cosmological model exhibits quintessence behavior (for CC data set) but for joint (CC+ Pantheon) data set, this model exhibits quintessence behavior at present $(z=0)$ and exhibits Pantheon behaviour in the late time. 
\subsection{Universe's Age}\label{sec:6.4}
A cosmological model's cosmic age $t(z)$, can be calculated in terms of $\mathit{z}$ in~\cite{tong2009cosmic} as
\begin{equation}{\label{31}}
t(z)= \int_{z}^{\infty} \frac{dz}{(1+z)H(z)} \,dz. 
\end{equation}
For the Hubble parameter $H(z)$ given in equation (\ref{21}) at the current time $(z = 0)$, we numerically determine the above integral and calculate the universe's current age. From now on, this model will have $t(z = 0) = 12.52$ Gyr (for CC data set) and $t(z = 0) = 12.96$ Gyr (for joint data set)~\cite{balbi2007lambda}.
\section{Conclusions}\label{sec:7}
In this study, the universe's behavior for a flat FLRW metric in $ f(R, L_{m})$  gravity is investigated. Subsequently, the motion equations for the flat FLRW space-time metric is derived. Additionally, the deceleration parameter's behavior has been examined. The deceleration parameter's evolution curve shows a recent change in the universe's phase from decelerated to accelerated. The present values of deceleration parameters are $q_{0}=-0.36$ (for CC data) and $q_{0}=-0.50$ (for joint data set) and  the transition red-shift corresponding to the best fit value of model parameters is  $ z_{t} = 0.747 $ (for CC data) and $ z_{t} = 0.72  $ (for joint data set). 
\vspace{0.3cm}\\
The obtained best fit values $h_{0}=66.6^{+1.1}_{-1.1} Km/(s.mpc)$, $w_{0}=-1.24^{+0.17}_{-0.17}, w_{1}=0.338^{+0.068}_{-0.075}, n=1.03^{0.17}_{-0.062}$ are compatible with the observational constraint values estimated for CC data in \cite{solanki2023cosmic}, \cite{hinshaw2013nine} and \cite{jaybhaye2022cosmology} 
respectively, whereas $h_{0}=68.9^{+1.9}_{-1.9} Km/(s.mpc)$, $w_{0}=-1.64^{+0.24}_{-0.24}, w_{1}=0.48^{+0.11}_{-0.11},$ \\ $n=1.02^{0.17}_{-0.079}$ are consistent with estimated in \cite{lalke2024cosmic}, \cite{hinshaw2013nine} and \cite{jaybhaye2022cosmology}
for joint data sets respectively.
\vspace{0.3cm}\\
The behaviour of physical parameters, viz. ($ \rho $ and $ p $) have been discussed for best-fit values of model parameters. The energy density of the considered model exhibits a positive behaviour. In contrast, the pressure has become negative from the recent past. 
The faster expansion of the cosmic cosmos at this late time may be caused by the pressure's navigate behavior.
 \vspace{0.3cm}\\
However, our overall observation is that the model is compatible with the present day accelerating universe observations. 
\vspace{0.3cm}\\
In addition, jerk and snap parameter have been discussed. Figure $ (\ref{fig:11})$ illustrated that at late time evaluation of the universe our model is similar to $\Lambda$ CDM model for both datasets viz. CC data and CC+Pantheon data. The Om diagnostic parameter's development profile demonstrates  that our $ f(R, L_{m})$ cosmological model exhibits quintessence behavior (for CC data) but for CC+ Pantheon data set, this model exhibits quintessence behavior at present$(z=0)$ and exhibits Pantheon behaviour in the late time. At last, we obtain the present universe's age is $t_{0} =12.52$ Gyr (for CC data set) and $t_{0} =12.96$ Gyr (for joint data set) for this model.

\section*{\textbf{Acknowledgements}}
Authors are thankful to the learned referee for the positive constrictive valuable suggestion. R. Garg expresses his gratitude to Dr. Ashutosh Singh for enlightening discussions. G. P. Singh and S. Ray are thankful to the Inter-University Centre for Astronomy and Astrophysics (IUCAA), Pune, India for providing the Visiting Associateship under which a part of this work was carried out. S. Ray also acknowledges the facilities under ICARD at CCASS, GLA University, Mathura, India.

\bibliographystyle{unsrt}
\bibliography{PAPER_WRITING}

\end{document}